# Real-time Flux Density Measurements of the 2011 Draconid Meteor Outburst

S. Molau (1,2), G. Barentsen (1,3)

(1) International Meteor Organization
(2) Abenstalstr. 13b, 84072 Seysdorf, Germany (sirko@molau.de),
(3) University of Hertfordshire, College Lane, Hatfield, AL10 9AB, U.K. (geert@barentsen.be)

Abstract

During the 2011 outburst of the Draconid meteor shower, members of the Video Meteor Network of the International Meteor Organization (IMO) provided, for the first time, fully automated flux density measurements in the optical domain. The data set revealed a primary maximum at 20:09 UT ±5 min on 8 October 2011 (195.036° solar longitude) with an equivalent meteoroid flux density of $(118 \pm 10) \times 10^{-3}$ km$^{-2}$ h$^{-1}$ at a meteor limiting magnitude of +6.5, which is thought to be caused by the 1900 dust trail of comet 21P/Giacobini-Zinner. We also find that the outburst had a full width at half maximum (FWHM) of 80 minutes, a mean radiant position of alpha=262.2°, δ=+56.2° (±1.3°) and geocentric velocity of $v_{geo}$=17.4 km/s (±0.5 km/s). Finally, our data set appears to be consistent with a small sub-maximum at 19:34 UT ±7 min (195.036° solar longitude) which has earlier been reported by radio observations and may be attributed to the 1907 dust trail.

## 1. Introduction

The International Meteor Organization (IMO) Video Meteor Network is a joint effort of amateur astronomers who obtain video observations of meteors on a regular basis using the MetRec software (Molau 2001, Molau 2013). The aim of the network is to understand the properties of meteor showers, by measuring the position, brightness and appearance time of large numbers of meteors at visible wavelengths.

The history of the network can be divided into three phases. Starting in 1999, the first stage was characterized by the setup of the network, software development and initial data collection. The second phase started in 2006, when the database contained single-station astrometry for more than a quarter million meteors. Comprehensive statistical analyses were started in this period to detect meteor showers and obtain their parameters automatically. The third stage from 2010 on saw the addition of flux density estimates.

The estimation of meteoroid flux densities is achieved by combining the known characteristics of each camera system, such as collection area and integration time, with an automated measurement of the limiting magnitude during each observing session. The incoming flux of meteoroids per unit time under normalized conditions is then calculated (the normalized conditions assume a meteor layer at 100 km altitude, 1000 km$^2$ effective collection area, known meteor shower velocity and population index, radiant at zenith, 6.5 mag absolute limiting magnitude for meteors). The flux density of meteor showers was first determined for the 2011 Lyrids (Molau et al. 2011) and subsequently for all major meteor showers. The data can be analysed and visualized by a web service called *MetRec flux viewer*.

Flux density graphs are normally not made available in in real-time, because a manual step of quality control is necessary to remove false detections (e.g. by clouds or airplanes). Thus, the flux data are normally uploaded manually by the observers to the central web service with a delay of several days.

## 2. Real-time flux density measurements

For the expected Draconid outburst on 8 October 2011, we prepared an experiment in the IMO video network. For the first time, the observing data were transfered automatically and in real-time to the central server. We postponed the manual quality check because chances were small that false detections would be recognized as a Draconid meteor. They would primarily dilute the sporadic flux measurement, which was irrelevant for this experiment.

Most cameras in the IMO network are stationary and not prepared for mobile expeditions with remote internet access to escape adverse weather conditions. Some stations do not have an internet connection at all. The IMO network covers such a large area, however, that at least a few stations were expected to enjoy clear skies and contibute flux dentity measurmens in real-time.

In October 2011, 38 observers with 66 camera systems participated in the IMO network. About twenty observers agreed to join the experiment with their camera systems and upload the flux density data in real-time. Both the MetRec software and the MetRec flux viewer web service were adapted on short notice to allow continuous export and import of data. New flux density graphs were automatically created by the webservice every two minutes. The graphs were copied to a cache webserver, because the software was not designed to handle large numbers of parallel requests from internet web clients.

Our experiment was planned and implemented in less that four weeks and had a proof-of-concept character. As the weather turned bad in central Europe just a few days before the outburst, most observers were facing poor observing conditions. In the end, however, there were four cameras in Germany (KLEMOI operated by B. Brinkmann) , Slovenia (ORION1 operated by J. Kac and ORION2 operated by M. Govedic) and Portugal (TEMPLAR3 operated by R. Goncalves) that collected data under mainly clear skies and uploaded them in real-time. Based on these data sets, the live flux density profile was calculated and published without manual interaction. An animation sequence of the development of the flux density graph during the night is available online (see http://www.imonet.org/draconids/), which shows that that the Draconids peak was immediately and clearly apparent on the real-time flux graph.

## 3. Detailed analysis results

In the days after the outburst, the data set grew significantly when all other observers uploaded their flux density data as well. In the end, data from 57 video cameras were available, enabling us to carry out a thorough analysis of the outburst.

Figure 1 shows a high resolution Draconid flux density profile of eight hours centered on the peak time. Because the data set from that time interval contained 1605 Draconid meteors, fluxes could be computed by binning the data into intervals of only five minute length. We find that the primary peak occured at 20:09 UT ±5 min (195.036° solar longitude) with an equivalent[1] flux density equal to $(118 \pm 10) \times 10^{-3}$ km$^{-2}$ h$^{-1}$ at a meteor limiting magnitude of +6.5 assuming a population index of 2.0.

Figure 1 also shows a possible sub-maximum at 19:34 UT ±7 min (195.011° solar longitude), which is consistent with the findings of Steyaert (2013), who analysed radio observations and found the main peak between 20:00 and 20:10 UT and an early sub-maximum between 19:30 and 19:35 UT.

Comparing our results with the prediction of Vaubaillon (2011), it is likely that the primary peak is explained by an encounter with the 1900 dust trail of comet 21P/Giacobini-Zinner (predicted peak time 20:01 UT) and the early sub-maximum by the 1907 dust trail (predicted peak time 19:26 UT). In both cases, the observed peak was late by eight minutes.

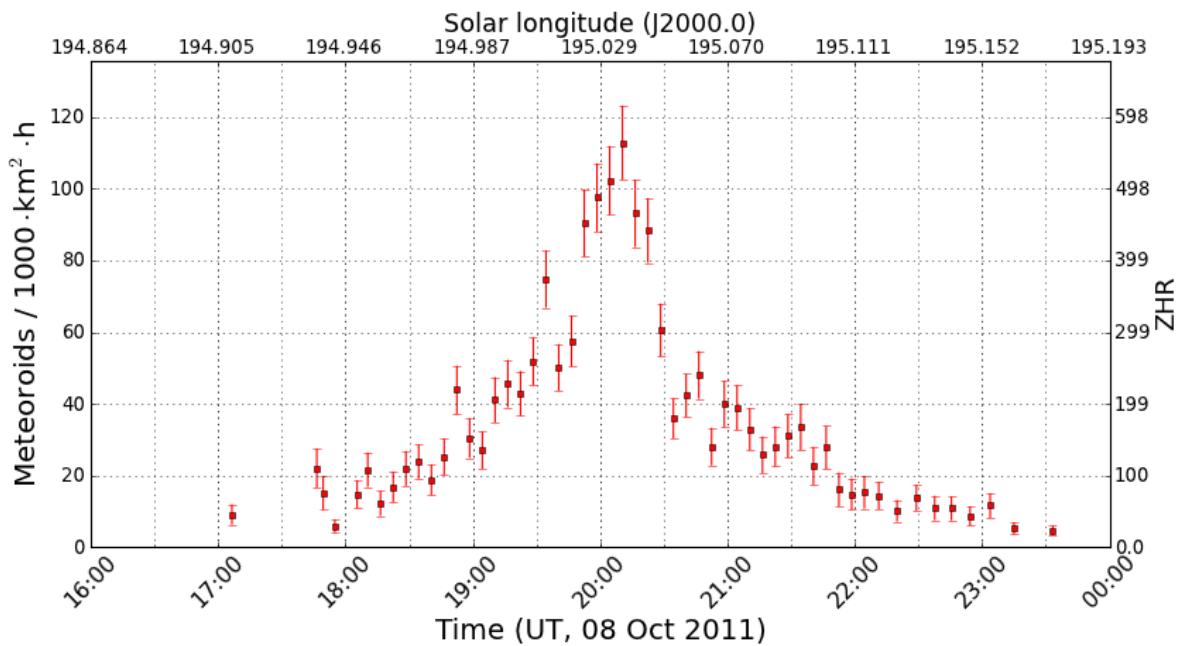

*Figure 1: High resolution flux density profile of the Draconids on October 8, 2011, derived from data of the IMO Video Meteor Network.*

To determine the full width at half maximum (FWHM) of the outburst, a third order polynomial was fitted independently to the ascending and descending branch of the activity profile (Figure 2). The FWHM was determined to be 80 minutes (19:20-20:40 UT) with the descending branch being clearly steeper than the ascending. Figure 2 includes also visual data from the IMO quick look analysis (http://www.imo.net/live/draconids2012). They yield a peak at 20:12 UT with almost identical FWHM. Peak visual equivalent[1] ZHR was about 300.

To determine the radiant position and meteor shower velocity, we analysed 2425 single-station Draconids recorded by IMO video network cameras on October 8/9, 2011. This data set was larger because it included also stations without limiting magnitude information. We applied the same single-station method as in previous meteor shower analyses (Molau 2008) including compensation of the zenith attraction. The resolution was chosen higher than usual (0.1° in right ascension and declination, and 0.1 km/s in velocity). We obtained a mean radiant at alpha=262.2°, delta=+56.2° (±1.3°) and a velocity of $v_{geo}$=17.4 km/s (±0.5 km/s). The radiant position is in good agreement with the result of other observers (e.g. Segon et al 2013), but the velocity is significantly smaller. We believe that our velocity is underestimated, because velocity determination from single station observations relies on an assumption of the (average) meteor altitude, which can be precisely measured in case of double-station observations.

---
[1] We use the term *equivalent* to underline that the peak flux density and ZHR was reached in a time interval of five minutes length, not a full hour.

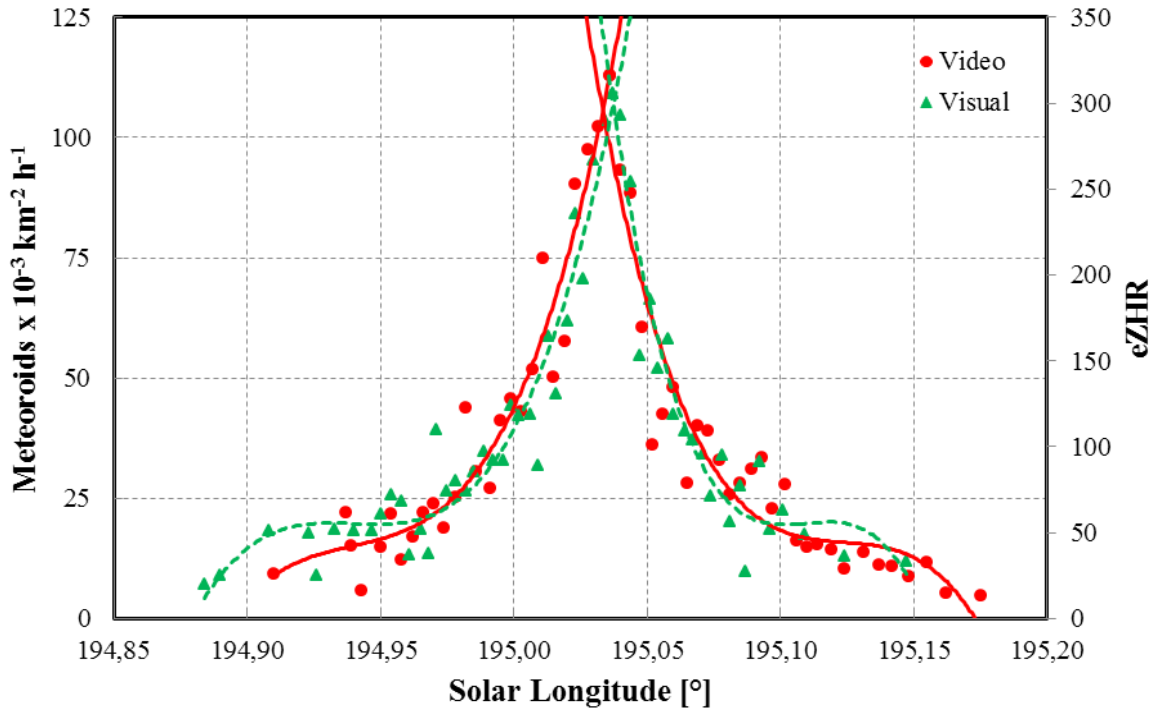

*Figure 2: Comparison of video observation (dots and solid line) and visual data (triangles and dashed line) of the 2011 Draconid outburst. The lines represent a third order polynomial fit to the ascending and descending activity branches, respectively.*

### 4. Summary

During the Draconid meteor outburst on 8 October 2011, members of the IMO Video Meteor Network successfully obtained, for the first time, automated flux density measurements in the optical domain. The subsequent offline analysis revealed a mean radiant position that agrees well with the results of other teams, albeit with a lower velocity. The main peak that can be attributed to the 1900 dust trail occured at 20:09 UT, and we confirm an early sub-maximum at 19:34 UT, which may be associated with the 1907 dust trail. The peak flux density was (118 ± 10) x $10^{-3}$ $km^{-2}$ $h^{-1}$ based on a population index of 2.0, and the full width at half maximum was 80 minutes.